\documentclass[journal]{IEEEtran}
\usepackage{cite}
\ifCLASSINFOpdf
   \usepackage[pdftex]{graphicx}
\else
\fi
\usepackage{amsmath}
\usepackage{url}
\begin{document}

\title{Bare Demo of IEEEtran.cls\\ for IEEE Journals}

\title{Direct-Write Ion Beam Irradiated Josephson Junctions}

\author{Ethan~Y.~Cho,~\IEEEmembership{Member,~IEEE,}
Hao~Li,~\IEEEmembership{Member,~IEEE,}
and Shane~A.~Cybart,~\IEEEmembership{Member,~IEEE,}
\thanks{E. Y. Cho, H. Li and S. A. Cybart are with the Department of Electrical and Computer Engineering, University of California, Riverside, Riverside, CA, 92093 USA email:eycho@ucr.edu}
\thanks{This
work was supported in part by the Air Force Office of Scientific Research under Grant FA955015-1-0218, in part by the National Science Foundation under Grant 1664446, in part by the National Institutes of Health under Contract No. j1R43EB023147-01, in part by the University of California Office of the President, Multi-campus Research Programs and Initiatives under Award No. 009556-002, and in part by the Army Research Office Grant W911NF1710504.}
}

\maketitle

\begin{abstract}

We highlight the reproducibility and level of control over the electrical properties of YBa$_2$Cu$_3$O$_7$ Josephson junctions fabricated with irradiation from a focused helium ion beam.
Specifically we show the results of electrical transport properties for several junctions fabricated using a large range of irradiation doses.At the lower end of this range, junctions exhibit superconductor-normal metal-superconductor (SNS) Josephson junction properties.
However, as dose increases there is a transition to electrical characteristics consistent with superconductor-insulator-superconductor (SIS) junctions. To investigate the uniformity of large numbers of helium ion Josephson junctions we fabricate arrays of both SNS and SIS Josephson junctions containing 20 connected in series. Electrical transport properties for these arrays reveal very uniform junctions with no appreciable spread in critical current or resistance.
\end{abstract}


\section{Introduction}

Since the discovery of high-transition temperature superconductors (HTS) there has been a great deal of progress in the development of Josephson junctions from these complex anisotropic materials \cite{mitchell20162d,pawlowski2018static,taylor2017hts,adachi2016fabrication,muller2019josephson}.
One particular approach to device fabrication that is recently gaining traction is to utilize irradiation to create a narrow nanoscale region of crystalline defects within the material that can be utilized as a Josephson junction.
This process was developed many decades ago; initially utilizing electron beam irradiation \cite{booij1997proximity} and later ions \cite{white1988controllable,Tinchev1990investigation}.
Historically, these types of Josephson junctions featured low critical voltages ($I_\text{C}R_\text{N}$) and large excess currents ($I_\chi$) \cite{cybart2005planar} due to very weak diffusive Josephson "barriers" from device feature resolution limits, set by the particular process (typically tens of nanometers) \cite{bergeal2007using,cybart2008series}. 
 Recently, our group demonstrated that by confining the irradiated region to sub-10-nm dimensions with a finely focused helium ion beam (FHIB) that we can dramatically affect the electrical properties of ion irradiated Josephson devices.
 Specifically, ion irradiated junctions were demonstrated with insulating tunnel barriers, no excess current, and very high resistance ($R_\text{N}$) \cite{cho2018superconducting,cybart2015nano,cho2015yba2cu3o7,cho2018direct}. 
 
\begin{figure}
    \centering
    \includegraphics[width=\linewidth]{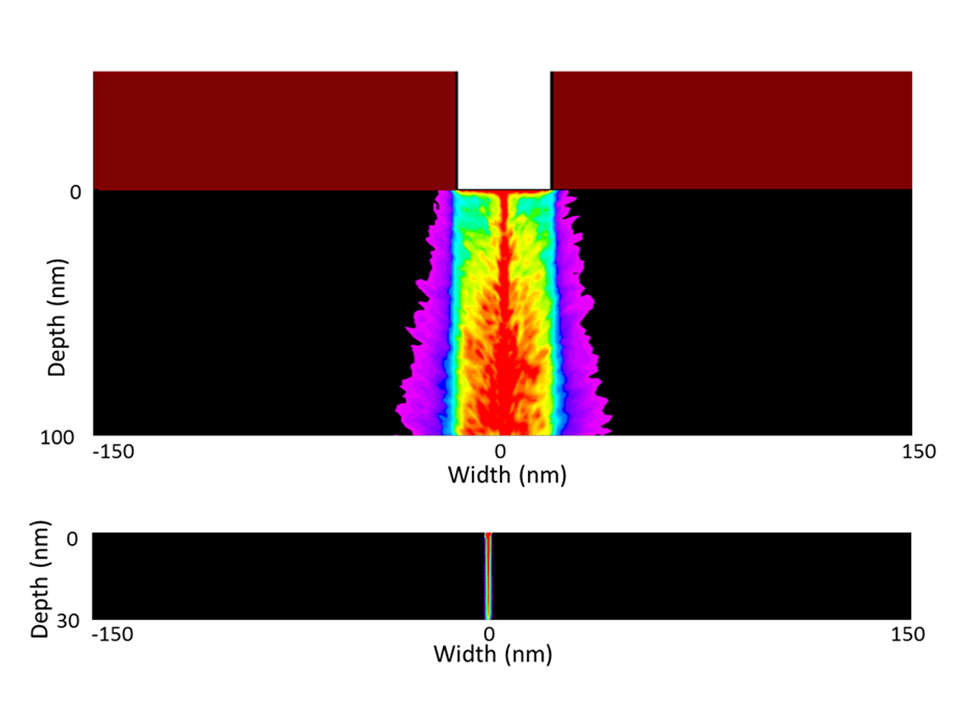}
    \caption{Comparison of disordered region in a 200~nm-thick YBCO by 200~keV neon broad ion beam irradiation (top) and 25~nm-thick YBCO irradiated by 30~keV helium focused ion beam (bottom).}
    \label{irradiation}
\end{figure}

 To illustrate the difference between our prior-art devices and current helium ion Josephson junctions (HI-JJ) we compare Monte Carlo ion implantation simulations performed with Silvaco Athena software of a device fabricated using 200~keV neon implantation through a nanofabricated mask \cite{cybart2009very} with that of a HI-JJ \cite{cybart2015nano}.  The result is shown in  Fig. \ref{irradiation} as a cross section of the devices cut parallel to the direction of the current. The disordered region of the HI-JJ is over an order of magnitude narrower than that of the masked device, illustrating the fine resolution afforded by the gas field ion source. We note that the thickness of HI-JJs are currently limited to tens of nanometers due to the maximum operating voltage of commercial GFIS sources $\sim$35~kV. 

In contrast to masked ion beam junctions, in the HI-JJ process, disorder is only induced in a very narrow region with a density near the atomic density of the material. In this paper, we discuss how this fine feature control allows for insulating barrier devices that exhibit higher resistances and significantly less excess current. Furthermore, we demonstrate control over junction parameters and show the reproducibility and uniformity of HI-JJ with  measurements of both superconductor-insulator-superconductor (SIS) and superconductor-normal metal-superconductor (SNS) YBa$_2$Cu$_3$O$_{7-\delta}$ (YBCO) junction arrays.

\section{Experimental}
\subsection{Superconductor-insulator transition}
The quality of Josephson junctions is governed by the $I_\text{C}R_\text{N}$, and should be maximized for most applications. The optimum ratio between $I_\text{C}$ and $R_\text{N}$ varies per application  but higher $I_\text{C}R_\text{N}$ represent better quality.
For a tunneling process, $R_\text{N}$ is related to the barrier potential. Whereas $I_\text{C}$ is related to the barrier width and the critical current density of the material. 
To maximize $R_\text{N}$ the junction barrier needs have very high potential while maximizing $I_\text{C}R_\text{N}$ requires the barrier to be as thin as possible. The FHIB can create disorder with very little straggle ($\sim$3~nm) to achieve these goals.

\begin{figure}
    \centering
    \includegraphics[width=\linewidth]{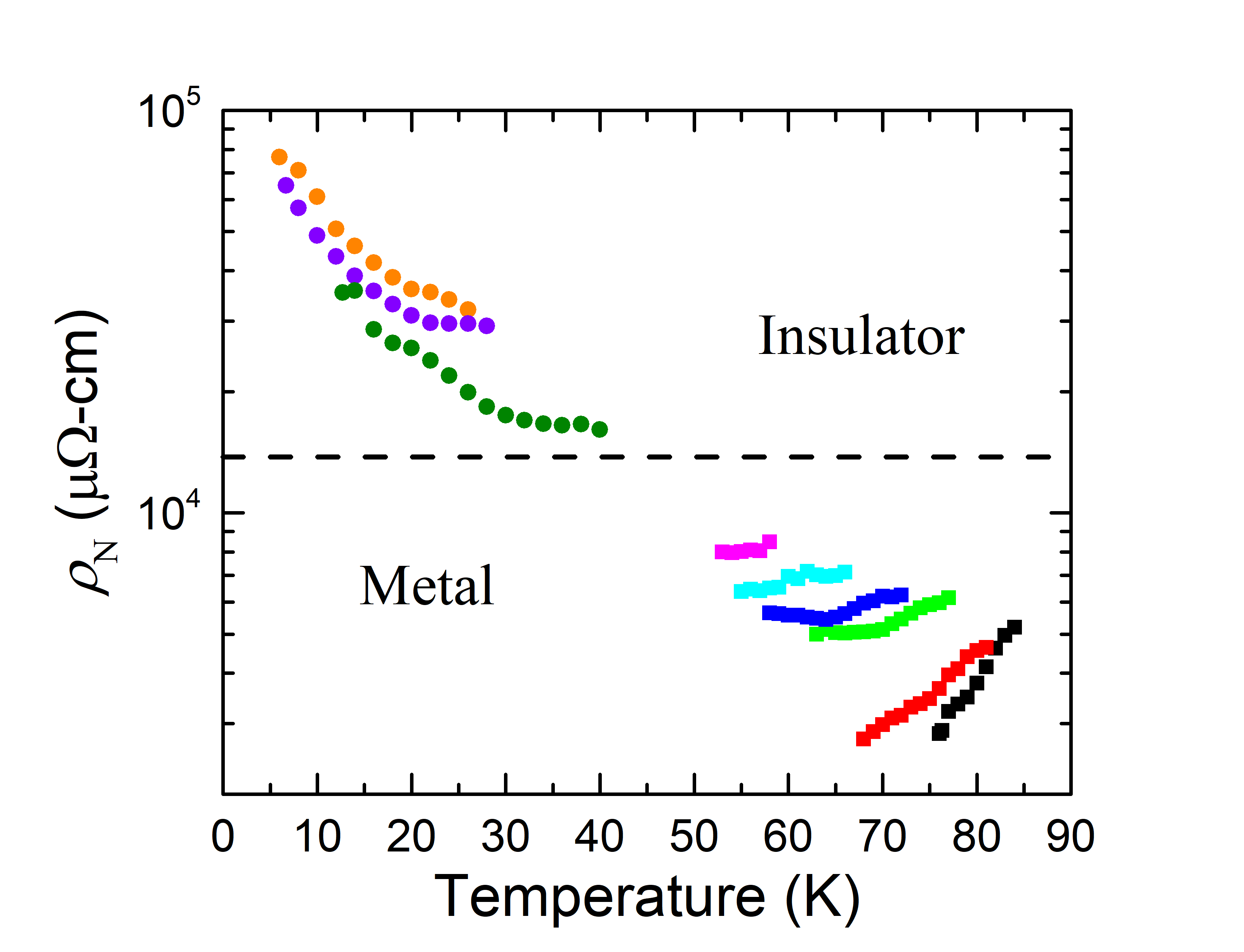}
    \caption{The resistivity of junction barrier irradiated with fluence of $1.5, 2.0, 2.5, 3.0, 3.5, 4.0, 6.5, 7.0, \text{and}~8.5\times10^{16}$~ions/cm$^2$ where full circles represent insulating behavior and full squares represent metallic behavior.}
    \label{fig:rhoMIT}
\end{figure}

To analyze the control over device properties we investigate the junction barrier resistivity as functions of temperature and ion irradiation dose. For this work 9 HI-JJ junctions were fabricated on the same chip using doses of $1.5, 2.0, 2.5, 3.0, 3.5, 4.0, 6.5, 7.0, \text{and}~8.5\times10^{16}$~ions/cm$^2$. Current-voltage (\textit{I-V}) characteristics were measured for each device over a large range of temperature and fit with the Stewart-McCumber model to determine the $I_\text{C}$ and voltage state resistance, $R_\text{N}$. $R_\text{N}$ is converted to resistivity, $\rho_\text{N}$ by using the lithographically defined junction width (4~$\mu$m) and taking the length of the barrier (in the direction of the current) to be 3~nm, based on our simulations. The resulting $\rho_\text{N}(T)$ is plotted in  Fig. \ref{fig:rhoMIT}. At the lowest doses $\rho_\text{N}(T)$ rapidly decreases with decreasing temperature like a metal while at the highest doses it increases like an insulator. At doses near $4\times10^{16}$~ions/cm$^2$ $\rho_\text{N}(T)$ is independent of temperature and constant. Although there has between some individual reports of HTS junctions with insulating barriers\cite{yi1996tunneling} only the HI-JJ process has been able to demonstrate a continuous transition from SNS to SIS junctions\cite{cho2015yba2cu3o7}. Additionally, the $\rho_\text{N}(T)$ of the junction barriers from FHIB irradiation agrees very well with $\rho(T)$ obtained from bulk film and broad beam irradiation \cite{lang2006ion}. In our prior work we have confirmed the presence of quasiparticle tunneling in the insulating devices \cite{cybart2015nano} however we remark that the increase in $\rho_\text{N}(T)$ signals the presence of an additional transport mechanism through the barrier because tunneling is a temperature independent process. We speculate this to be a form of hopping conduction. 

\begin{figure}
    \centering
    \includegraphics[width=\linewidth]{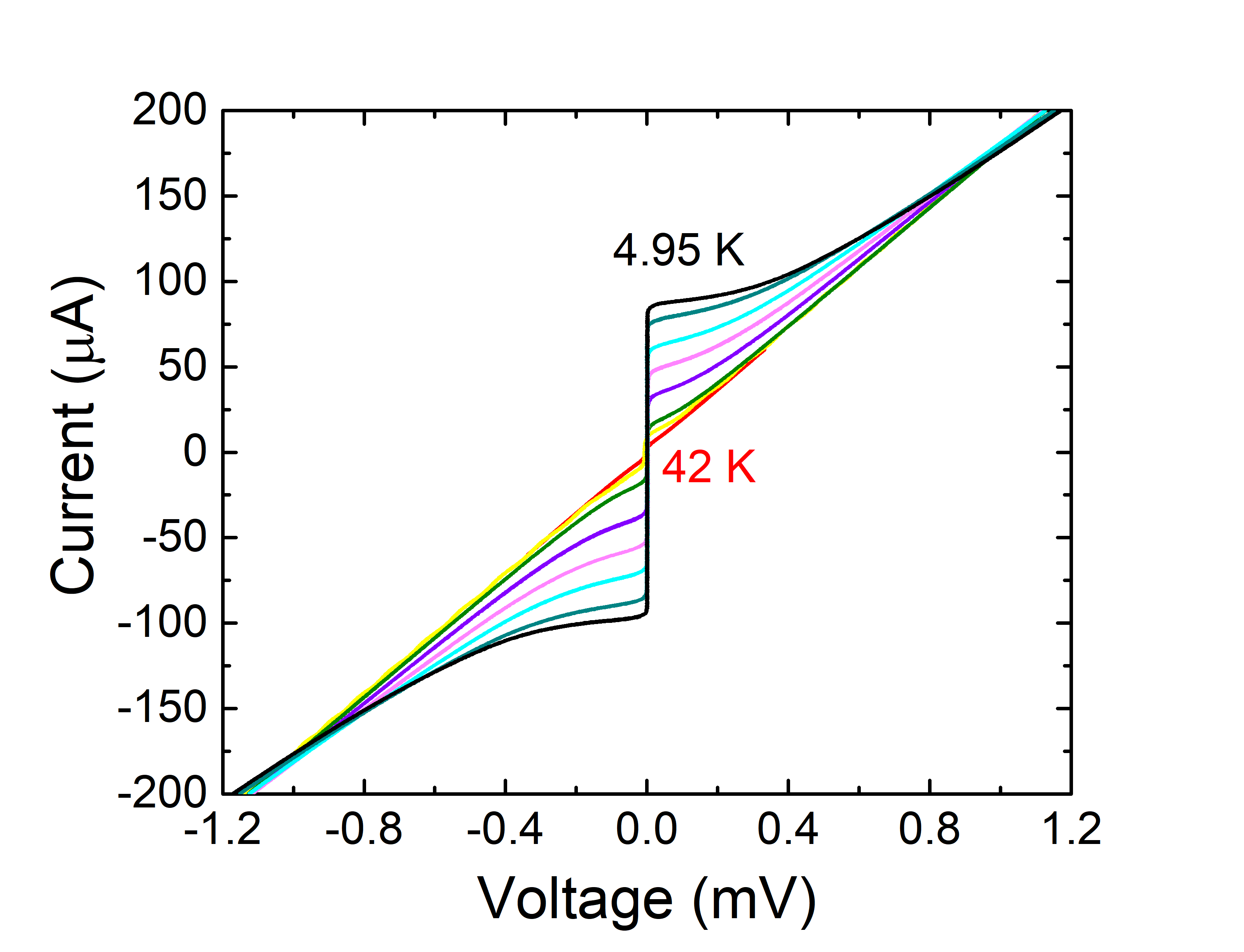}
    \caption{Current-voltage characteristics of a 4~$\mu$m-wide junction with a constant $R \approx 5.6~\Omega$ over 40~K.}
    \label{fig:bestIV}
\end{figure}

In Fig. \ref{fig:bestIV} we show the \textit{I-V} of a 4~$\mu$m-wide junction fabricated with a dosage in the middle of the superconductor insulator transition with $R_\text{N}$ = 5.6~$\Omega$ over a wide temperature range without any excess current.

\subsection{HI-JJ Arrays}
\begin{figure}
    \centering
    \includegraphics[width=\linewidth]{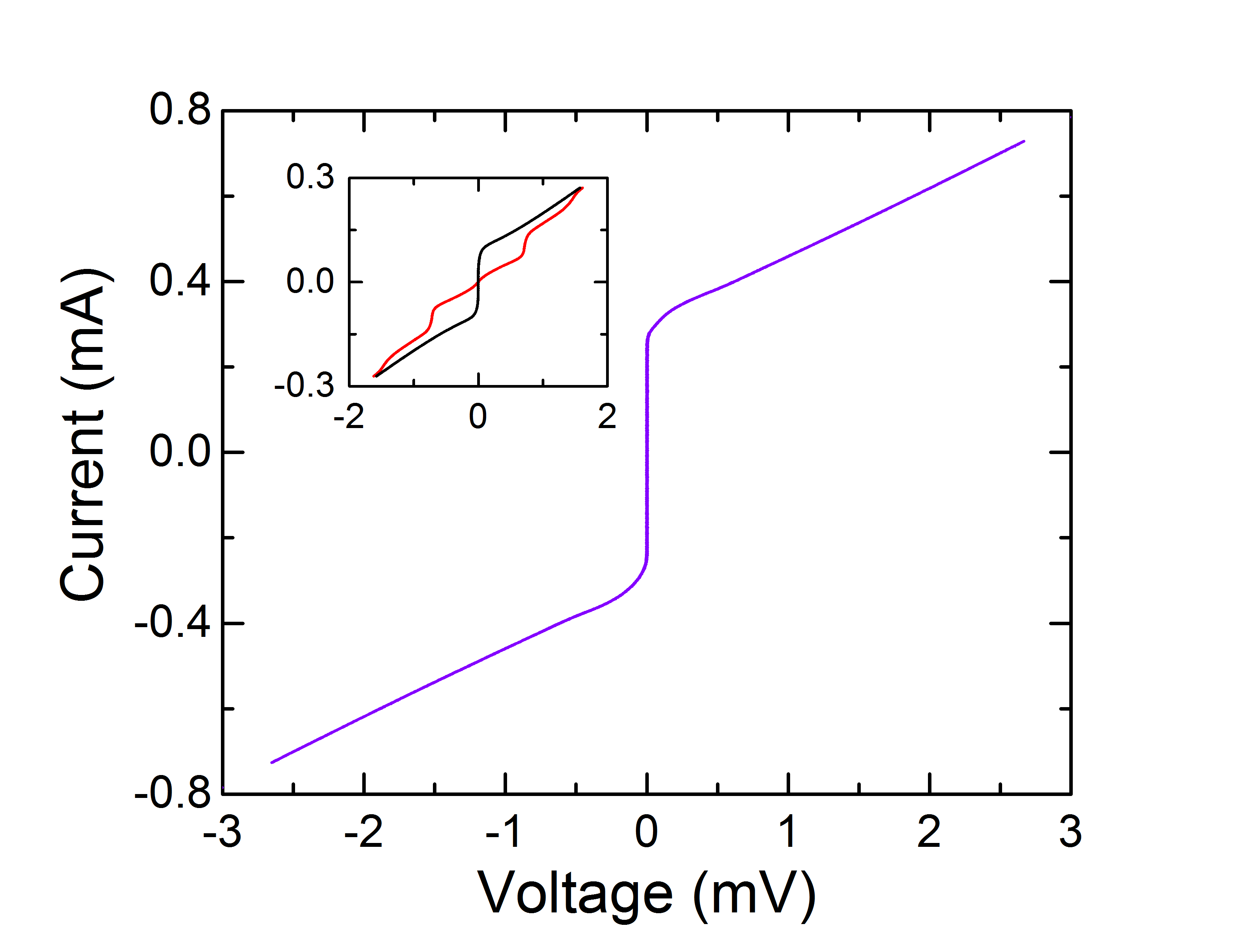}
    \caption{20-Josephson junctions in series with uniform junction parameters. The inset shows a giant Shapiro step irradiated with 17~GHz RF.}
    \label{fig:JJarray}
\end{figure}

To study the reproducibility we made fabricated series arrays containing 20 HI-JJ spaced 1~$\mu$m apart with irradiation doses suitable for SIS and SIS junctions. 
Data for the SNS array measured at 55~K are shown in Fig.~\ref{fig:JJarray}. The \textit{I-V} is very uniform and shows RSJ-like properties. Very little rounding is observed near the critical current which is typically associated with multi-junction arrays. To test the uniformity of $R_\text{N}$ the array was irradiated with 17~GHz RF and giant Shapiro steps were observed at 20 times the voltage expected for that of a single junction Fig.~\ref{fig:JJarray}(inset). This indicated that the quantized constant voltages are occurring at the same bias current confirming no appreciable spread in $R_\text{N}$. 

\begin{figure}
    \centering
    \includegraphics[width=\linewidth]{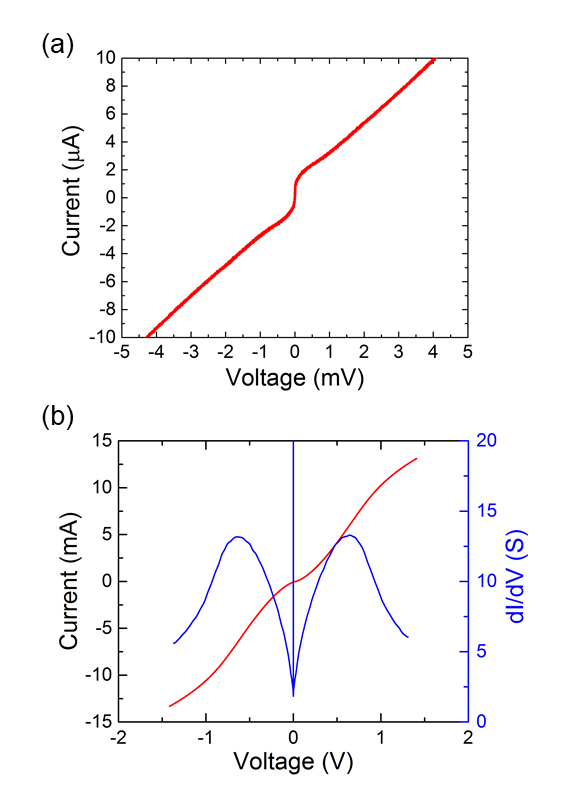}
    \caption{(a) \textit{I-V} of an array of 20 SIS junctions in series at 4~K with $R~\sim~500~\Omega$ and $I_\text{C}$ less than 2~$\mu$A. (b) \textit{I-V} of the same array as (a) biased beyond the gap voltage (red) with a peak at 660~mV, and the dynamic conductance of the array measured using lock-in amplifier (blue).}
    \label{fig:SIS}
\end{figure}

A series array of 20 SIS junctions was fabricated using a higher dose of irradiation. and Fig.~\ref{fig:SIS}(a) shows its \textit{I-V} measured at 4~K.  Like in the case of the SNS array the characteristics are smooth and resemble that of a single junction. However, we note that the resistance is 20 times higher than that of similarly fabricated  single junction devices \cite{cybart2015nano}. Due to limitations of our 4~K measurement setup we were unable to perform RF measurements for this chip to investigate Shapiro steps. However, because these are insulating barrier junctions we were able to investigate the quasiparticle tunneling spectrum. Using a lock-in amplifier the dynamic conductance of the array was measured at high voltage bias and the result is shown in Fig.~\ref{fig:SIS}(b). The large spike in the center is from the Josephson supercurrent and the peaks near 660~mV are attributed to the YBCO energy gap of all of the junctions. The energy gap of single HI-JJ using the same wafer of YBCO was previously measured to be 33~mV\cite{cybart2015nano}.
Here the array shows the same peak at 20 times the voltage assuring us that all 20 junctions are present with well formed insulating barriers.

\section{Conclusion}
We have shown that by tuning the disorder density in HI-JJ it is possible to control the electrical properties of the junctions and create both SNS and SIS type junctions in YBCO. The helium ion beam offers great control over junction properties and uniformity which broadens the applications of ion irradiated junctions. Furthermore, direct writing of YBCO can also be utilized to combine junctions with nanowires \cite{cho2018superconducting} and nanostructured SQUIDs \cite{cho2018direct}. 
HI-JJ fabrication requires less process steps than prior-art techniques and can  easily be scaled to a wafer process. With further process refinements HI-JJ could bring many applications of superconductive electronics out of the research lab and into the commercial sector.

\section*{Acknowledgment}

The authors would like to acknowledge all the students in the lab for their passion about HTS electronics.

\bibliographystyle{IEEEtran}
\bibliography{main}
\end{document}